\documentclass[preprint,12pt]{aastex}

\begin{document}

\newcommand\tna{\,\tablenotemark{a}}
\newcommand\tnb{\,\tablenotemark{b}}

\title{The Mid-Infrared Instrument for the James Webb Space Telescope, I: Introduction}

\author{G. H. Rieke\altaffilmark{1}, G. S. Wright\altaffilmark{2}, T. B\"oker\altaffilmark{3}, 
J. Bouwman\altaffilmark{4}, L. Colina\altaffilmark{5}, Alistair Glasse\altaffilmark{2},  K. D. Gordon\altaffilmark{6,7},  T. P. Greene\altaffilmark{8}, Manuel G\"udel\altaffilmark{9, 10}, Th. Henning\altaffilmark{4}, K. Justtanont\altaffilmark{11}, P.-O. Lagage\altaffilmark{12}, M. E. Meixner\altaffilmark{6,13}, H.-U. N\o rgaard-Nielsen\altaffilmark{14}, T. P. Ray\altaffilmark{15}, M. E. Ressler\altaffilmark{16},  E. F. van Dishoeck\altaffilmark{17},  \& C. Waelkens\altaffilmark{18}.}

\altaffiltext{1}{Steward Observatory, 933 N. Cherry Ave, University of Arizona, Tucson, AZ 85721, USA}
\altaffiltext{2}{UK Astronomy Technology Centre, Royal Observatory, Edinburgh, Blackford Hill, Edinburgh
EH9 3HJ, UK}
\altaffiltext{3}{European Space Agency, c/o STScI, 3700 San Martin Drive, Batimore, MD 21218, USA}
\altaffiltext{4}{Max-Planck-Institut f\"ur Astronomie, K\"onigstuhl 17, D-69117 Heidelberg, Germany}
\altaffiltext{5}{Centro de Astrobiolog\'ia (INTA-CSIC), Dpto Astrof\'isica, Carretera de Ajalvir, km 4, 28850 Torrej\'on de Ardoz, Madrid, Spain}
\altaffiltext{6}{Space Telescope Science Institute, 3700 San Martin Drive, Baltimore, MD 21218, USA}
\altaffiltext{7}{Sterrenkundig Observatorium, Universiteit Gent, Gent, Belgium}
\altaffiltext{8}{Ames Research Center, M.S. 245-6, Moffett Field, CA 94035, USA}
\altaffiltext{9}{Dept. of Astrophysics, Univ. of Vienna, T\"urkenschanzstr 17, A-1180 Vienna, Austria}
\altaffiltext{10}{ETH Zurich, Institute for Astronomy, Wolfgang-Pauli-Str. 27, CH-8093 Zurich, Switzerland}
\altaffiltext{11}{Chalmers University of Technology, Onsala Space Observatory, S-439 92 Onsala, Sweden}
\altaffiltext{12}{Laboratoire AIM Paris-Saclay, CEA-IRFU/SAp, CNRS, Universit\'e Paris Diderot, F-91191 Gif-sur-Yvette, France}
\altaffiltext{13}{The Johns Hopkins University, Department of Physics and Astronomy, 366 Bloomberg Center, 3400 N. Charles Street, Baltimore, MD 21218, USA}
\altaffiltext{14}{National Space Institute (DTU Space), Technical University of Denmark, Juliane Mariesvej 30,
DK-2100 Copenhagen, Denmark}
\altaffiltext{15}{Dublin Institute for Advanced Studies, School of Cosmic Physics, 31 Fitzwilliam Place, Dublin 2, Ireland}
\altaffiltext{16}{Jet Propulsion Laboratory, California Institute of Technology, 4800 Oak Grove Dr. Pasadena, CA 91109, USA}
\altaffiltext{17}{Leiden Observatory, Leiden University, PO Box 9513, NL-2300 RA Leiden, the Netherlands}
\altaffiltext{18}{Institute of Astronomy KU Leuven, Celestijnenlaan 200D,3001 Leuven, Belgium}
\altaffiltext{}{}
\altaffiltext{}{}
\altaffiltext{}{}

\begin{abstract}

MIRI (the Mid-Infrared Instrument for the James Webb Space Telescope (JWST)) 
operates from 5 to 28.5 $\mu$m and combines over this range: 1.) 
unprecedented sensitivity levels; 2.) sub-arcsec angular resolution;  3.)
freedom from atmospheric interference; 4.) the inherent 
stability of observing in space; and 5.) a suite of versatile capabilities including imaging, low 
and medium resolution spectroscopy (with an integral field unit), and 
coronagraphy. We illustrate the potential uses of this unique combination of
capabilities with various science examples: 1.) imaging exoplanets; 2.) transit
and eclipse spectroscopy of exoplanets; 3.) probing the first stages of star and
planet formation, including identifying bioactive molecules; 4.) determining
star formation rates and mass growth as galaxies are assembled;
and 5.) characterizing the youngest massive galaxies. 

\end{abstract}


\section{Introduction}

The growth in capabilities for infrared astronomy since the 1960s \citep{low2007} 
has been spectacular. What is little-appreciated is that the gains are not 
evenly distributed over even the wavelength range accessible from the 
ground. In the near infrared, there has been a gain of about 10,000 in 
detection limits and instruments have gone from single detectors to mosaics 
of arrays totaling nearly 100 million pixels \citep{caldwell2004, dalton2006}. 
By comparison, the sensitivity gains for mid-IR photometry 
from the 1970s are more like a factor of 30 (mostly due to the growth in 
telescope size) and until recently the largest arrays in use have no more than 100 thousand 
pixels \citep{kataza2000, lagage2000}\footnote{VISIR has recently been upgraded to a a million pixels}, that is 1000 times fewer than the most ambitious instruments in the 
NIR. Some major telescopes (e.g., Gemini, Keck) do not even offer 
general-purpose mid-IR instruments, i.e. wide field imagers and moderate 
resolution spectrometers. 

The reason for this disparity is that the huge thermal background from the 
atmosphere and groundbased telescopes both blinds mid-infrared instruments 
and also overwhelms large-format detector arrays. Consequently, the 
mid-infrared range has been strongly dependent on cooled telescopes in 
space such as IRAS (the Infrared Astronomical Satellite), ISO (the Infrared Space 
Observatory), {\it Spitzer}, {\it Akari}, and WISE (the Wide-Field Infrared Survey Explorer). The elimination of thermal emission by the 
atmosphere and telescope for these missions resulted in huge gains in 
sensitivity. However, launching these telescopes into space imposed another 
limitation: they all had small apertures, of only 40 - 85 cm, and therefore 
provided limited angular resolution. In fact, for some of the mid-infrared 
bands on {\it Spitzer}, the ultimate sensitivity limit in deep imaging was set by 
confusion noise because of the limited angular resolution. 

A simple comparison indicates the advance afforded by the James Webb Sace Telescope (JWST). Compared with 
8-m groundbased telescopes (e.g., COMICS (the Cooled Mid-IR Camera and Spectrometer) 
on Subaru, CanariCam on the Gran Telescopio Canarias, VISIR (the VLT Imager and Spectrometer
for the mid-IR) on the VLT (Very Large Telescope)), the angular 
resolution is similar but with a potential gain in imaging sensitivity by a factor of about 3000 and for moderate 
resolution (R = $\lambda$/$\Delta \lambda $ $\sim$ 3000) 
spectroscopy a factor of about 1000 (see \citet{glasse2014} (Paper IX) for more information
about the MIRI sensitivity). Compared with {\it Spitzer}, the potential gain in the 
mid-IR where JWST is natural background limited (roughly 5 - 12 $\mu $m) is 
about a factor of 50 (see http://www.stsci.edu/jwst/science/sensitivity for 
more comparisons). The factor of 
seven improvement over that mission in the diffraction limit with JWST (e.g., from an image full
width at half maximum of 5\farcs8 at 24 $\mu$m with {\it Spitzer} to 0\farcs7 at 21 $\mu$m with JWST) 
is equivalent to going 
from a 330 Kpixel to a 16 Mpixel digital camera. That is, for the first time in the mid-IR, MIRI on JWST will {\it combine}: 1.) 
the incredible sensitivity levels just mentioned; 2.) sub-arcsec angular resolution;  
3.) freedom from atmospheric interference; 4.) the inherent 
stability of observing in space; and 5.) a suite of versatile capabilities including imaging, low 
and moderate resolution spectroscopy (with an integral field unit), and 
coronagraphy. 

This paper introduces a set of ten papers covering all aspects of MIRI and published by the Publications of the 
Astronomical Society of the Pacific, PASP volume 127, pages 584 - 696. The other papers can be consulted from PASP,
or at the MIRI web site\footnote{http://ircamera.as.arizona.edu/MIRI}.  
The latter has additional information about 
the instrument and will be updated as we progress toward launch in 2018.


\section{MIRI and the JWST}

``HST and Beyond" \citep{dressler1996} provided the initial momentum for a 
large, infrared-optimized space telescope\footnote{a concept then termed the Next 
Generation Space Telescope (NGST)} that has led to JWST. The report described 
the science opportunities for wavelengths between 1 and 5$\mu $m, strongly 
emphasizing the potential for extremely high redshift galaxy studies. Out of 
this concept came the ``First Light Machine" theme. 
The telescope was then envisioned as an ambitious but relatively inexpensive 
mission for which a focused science program was appropriate. The core 
instruments were aligned with the ``HST and Beyond" program; they consisted 
of an imager and a spectrometer both operating out to 5$\mu $m.  However, this report 
added ``Extension of this telescope's wavelength range shortward to about 
0.5$\mu$m and longward to at least 20$\mu$m would greatly increase its 
versatility and productivity. The Committee strongly recommends this course 
...."

The astronomy community understood the huge advances possible at the longer 
wavelengths, and almost immediately plans were explored for a Mid-Infrared 
Instrument (MIRI), building on the scientific momentum gained with the ISO results 
and the technical advances in preparation for {\it Spitzer}. The project decided that this instrument would 
require a 50/50 partnership between Europe and the US. Following separate 
competitive processes the MIRI team was selected on both sides of the 
Atlantic. Initially a cryostat to cool the instrument was to be provided by the European Space Agency 
(ESA), but this component was shifted to the US as part of the negotiations 
for ESA to supply the JWST launch vehicle. Eventually to save weight, a 
closed cycle cooler was adopted in place of the cryostat. 

The MIRI optical system was built by the European Consortium of national 
efforts, from Belgium, Denmark, France, Germany, Ireland, the Netherlands, 
Spain, Sweden, Switzerland, and the United Kingdom, led by Gillian Wright, 
the European Principal Investigator, and Alistair Glasse, Instrument 
Scientist. EADS-Astrium provided the project office and management, and the full 
instrument test was conducted at Rutherford Appleton Laboratory. The Jet Propulsion 
Laboratory (JPL) provided the core instrument flight software, the detector 
system, including infrared detector arrays obtained from Raytheon Vision 
Systems, collaborated with Northrop Grumman Aerospace Systems on the cooler 
development and test, and managed the US effort. The JPL Instrument 
Scientist is Michael Ressler and the MIRI Science Team Lead is George Rieke. The fully tested instrument 
was delivered to Goddard Space Flight Center at the end of May, 2012, where 
it is being integrated with the other components of JWST. The Space Telescope Science Institute (STScI) is the JWST operations center
and is developing software for the MIRI operations, user
interaction, data reduction, and archive. 
However, the success in building MIRI and the work to integrate it with the 
rest of JWST, to devise plans for its use, and to support that use with data 
pipelines and analysis tools depends on the efforts of many MIRI team 
members in both Europe and the US. Some of these people appear as co-authors 
of the following papers, but there are many more.

The MIRI design was driven by the need to provide a full suite of capabilities in a single instrument for wavelengths 
beyond 5 $\mu$m. It therefore combines imaging, coronagraphy, low 
resolution spectroscopy, and medium-resolution,  integral-field-unit 
spectroscopy.  Tables 1, 2, and 3 and Figure 1 provide a high level summary of the instrument capabilities,
based on current estimates (which remain uncertain in some areas, such as the level of emission from
the observatory into the long wavelength MIRI bands). This information is provided here for convenience,
but a more detailed summary is provided by \citet{glasse2014} (Paper IX). The instrument images will be diffraction limited
at all wavelengths, so their full width at half maxima is given by the expression 0.035 $\times$ $\lambda$($\mu$m) arcsec.
Additional information about MIRI can be found at
http://ircamera.as.arizona.edu/MIRI and http://www.stsci.edu/jwst/instruments/miri/. Up-to-date
estimates of the instrument sensitivity and other performance parameters should be obtained
from these websites or from the relevant observing planning tools at STScI. 
We describe a few of the science programs enabled by these 
capabilities in the following section.

\section{Sample Science Programs}

The overall science capabilities of JWST are presented in \citet{gardner2006}. 
Rather than trying to condense that discussion, here we describe a 
small number of programs to highlight the impact of the unique capabilities 
provided by MIRI and in some cases to introduce new suggestions from
the astronomical community for JWST programs.

\subsection{Exoplanet Imaging}

The discovery and study of exoplanets is the greatest astronomy 
initiative of the past decade. However, virtually all known exoplanets have 
been found indirectly - through their effect on the radial velocities of 
their stars, or by transit observations. Imaging of exoplanets is the next 
frontier; from the current successes (e.g., \citet{kalas2008, marois2008, lagrange2010, bailey2014}), 
it promises to be rich with discoveries. 
Exporing it thoroughly will reveal the regions of planetary systems 
beyond 10 AU and will return information needed to probe the origins and 
physical nature of the gas- and ice-giant planets that form in this region. 
The strength of the JWST instruments in general and MIRI in particular is their ability 
to image very faint planets, probing down to masses of order 0.1 - 0.2 
M$_{_{\mathrm{Jup}}}$.

Possible methods for direct imaging of exoplanets have been evaluated by \citet{beichman2010}. 
The MIRI coronagraph is optimized for such studies, 
since its four quadrant phase masks (4QPMs) allow information to be obtained 
to within $\sim$ $\lambda $/D of the star. The large telescope aperture places this 
limit at about 5 AU at 10 pc for the coronagraph channels at 11 - 16$\mu $m, 
and the sensitivity allows detection of low mass planets (see Figure 2). 
MIRI detects the thermal radiation (rather than reflected light) and hence 
is most sensitive to young exoplanets. Figure 2 summarizes the results of the 
\citet{beichman2010} 
study; it projects that many planets are most detectable with MIRI among the JWST instruments. For the 
25 most favorable cases, the study predicted that MIRI would have a high
success rate in finding any planets with average masses 
of 1--2 M$_{\mathrm{Jup}}$~at average separations of 60~AU. More 
importantly, JWST 
will achieve detections with multiple instruments, and these cases are the 
ones where the additional information will allow testing theories of exoplanet 
evolution.

\subsection{Transit and eclipse spectroscopy of exoplanets}

The stability of observations from space has allowed HST and {\it Spitzer} to make 
a series of remarkable observations of transits and secondary eclipses by 
exoplanets during the past decade, initiating the study of their physical 
properties such as energy balance and atmospheric composition and 
circulation. The expansion of this field of study with JWST is described in 
\citet{deming2009} and \citet{fortney2013}. From the former paper, we 
take one example that builds on the unprecedented sensitivity of MIRI and 
the stability of the space environment. 

Figure 3 shows the spectrum of a model atmosphere of a super-Earth with deep absorption 
around 15 $\mu $m due to CO$_{\mathrm{2}}$. The ratio of the signals in the two indicated
filters is a sensitive measure of the depth of the 15 $\mu$m absorption and hence of
the CO$_2$ content of the atmosphere of the planet; observations of 
secondary eclipses with the two filters would allow a 
sensitive search for CO$_{\mathrm{2}}$ absorption. The scheduled 
launch of the Transiting Exoplanet Survey Satellite (TESS) in 2017 will provide
optimal targets for transit and eclipse observations over the JWST mission.  Figure 4 
shows that a number of the anticipated super-Earths might be searched for CO$_2$ with a high priority program (the
signal to noise ratios are calculated assuming all eclipses available during the
JWST mission are observed). 

In addition, the figure demonstrates that high signal-to-noise could be achieved on
eclipses of many Neptune-sized planets, opening the possibility of obtaining
spectra of them. Taking the example of GJ 436b (a hot Neptune measured in transit with {\it Spitzer} photometry \citep{stevenson2010}), ten transits would yield a LRS spectrum with a ratio of signal
to noise of about 20 and a resolution of 100. Such a spectrum would allow detailed
studies of the atmospheric composition, including molecular features such as those from H$_2$O, NH$_3$, and
CH$_4$. Such species dominate the molecular atmospheric opacity where giant planets
have significant levels of thermal emission and hence are important in understanding
their energy balance and emission.  In general, the targets suitable for MIRI will 
also yield spectra with the JWST near infrared spectrometer (NIRSpec) that are sensitive to water and the 
CO$_{\mathrm{2}}$ band at 4.3 $\mu$m \citep{deming2009}. Monitoring 
planets around their orbits with these instruments will probe energy transport by their 
atmospheres as different regions of the planet are exposed and potentially, if the planet is on
an elliptical orbit, as it experiences differing levels if heating. Thus, MIRI will enable detailed 
characterization of super-Earths and larger planets. 

\subsection{Early stages of star and planet formation}

The embedded phase is a critical period in the
evolution of a young star, when its final mass and the
characteristics of any protoplanetary disk are determined.  Many
physical processes occur simultaneously in the first few $\times 10^5$
yr: infall in the collapsing envelope, formation of the disk, outflows
sweeping up and shocking the material, and UV photons heating and
dissociating the gas, thereby affecting the next generation of
protostars.  Large scale surveys of Galactic star-forming clouds with {\it
Herschel} and {\it Spitzer} have advanced our understanding enormously 
(Dunham et al. 2014), but studies of individual sources have
been hampered by poor spatial and spectral resolution and
sensitivity. MIRI’s integral field unit (IFU) spatial resolution down to 0\farcs2 and field of view (FOV) of
$3\farcs5-7''$ are well matched to the sizes of the young disks ($\sim$ 100
AU) and envelopes ($\sim$ 2000 AU). 

Figure 5 shows the {\it Spitzer-}IRS spectrum of a deeply embedded low-mass
protostar (b1c) and a less-obscured object for comparison (HH 46). The
mid-infrared continuum is generated by warm dust close to the
protostar in the inner envelope or young disk. Silicate grain
cores and ices in the cold surrounding envelope produce deep absorption features.  There are a few
windows (5--8 $\mu$m, near 14 $\mu$m and beyond 20 $\mu$m) for
looking deep into protostellar cores to characterize the physical
structure of the warm component (Cernicharo et al. 2000). These interstellar windows give access to a rich set of
spectral emission features that can trace chemical processes in the
youngest protoplanetary disks, including astrobiologically important
substances such as water and organic molecules. Ironically, much of 
the spectral range in these windows is blocked by the
terrestiral atmosphere.  MIRI will be able to
probe the earliest, most deeply embedded stages of star
formation, from very low-luminosity objects ($<0.1 L_\odot$, proto brown
dwarfs) to the highest luminosity ($> 10^5 L_\odot$), highly obscured
protostellar sources.

Figure 6 illustrates the richness of the molecular spectrum in a more
evolved protoplanetary disk around a T Tauri star for which the
envelope has dissipated (similar spectral richness is expected for
embedded protostars). The high spectral resolution provided by MIRI
will allow detailed study of the volatiles that dominate the
condensable mass and it will support increased understanding of
processes near snow lines during the period when volatile-rich
planetesimals form in gas-rich protoplanetary disks. The pioneering
spectroscopy of these regions with {\it Spitzer} (Carr \& Najita 2008,
Pontoppidan et al. 2010) has posed many questions that MIRI and JWST
with their combination of high sensitivity and spectral resolution can
answer: What are the abundances, temperatures, and distributions of
the disk molecules, and how do they differ between substellar and
higher mass young stars?  Are the gas and dust distributions
decoupled? The answers will give us a new perspective on the physical
and chemical conditions in the giant-planet forming zones of disks.


In general, JWST/MIRI is unique for many studies of the environments where stars and planets form. 
The fundamental vibrational transitions of many organic molecules are in the mid-IR, for example
CH$_4$ (7.7 $\mu$m), C$_2$H$_2$ (13.7 $\mu$m), HCN (14.0 $\mu$m), and CO$_2$ (15.0 $\mu$m). The first
two have no dipole moment and therefore are not accessible through rotational transitions in the mm-wave. 
H$_2$O has many transitions in the mid-IR, such as the bending mode at 6.0 $\mu$m. Ices of these materials
also absorb in broadened features near the fundamental transitions. The lowest transitions of H$_2$ also occur within
the MIRI spectral range, e.g., the J = 2 $-$ 0 S(0) line of para-H$_2$ at 28.22 $\mu$m and the J = 3 $-$ 1 (S(1) line of ortho-H$_2$ at 17.03 $\mu$m. Since the populations of these levels are usually thermalized, measurements of these
two lines provide a direct measure of the mass and temperature of the bulk of molecular gas at temperatures between
50 and 200K. Atomic fine structure lines throughout the mid-IR (e.g., [Ne II] 12.8 $\mu$m, [Ne III] 15.6 $\mu$m, [Ar II] 7.0 $\mu$m, [Ar III] 9.0 $\mu$m, \& [OIV] 25.91 $\mu$m, also [FeII] 25.988 $\mu$m) can probe the
hardness of the radiation field, the gas density,
metallicity, and the presence of shocks. This rich set of diagnostic lines can also explore the physics of the primeval jets and compare them with those from more evolved protostars. The high spatial resolution will let us identify source components, such as the outflow cavity walls, or internal disk shocks. The latter reflect the accretion history of the protostar and can test
theories of episodic versus steady-state accretion (Dunham et
al. 2014). These capabilities are complementary to those of ALMA (the Atacama Large Millimeter Array) for probing
the gas in forming stars and planets: the two capabilities have similar angular resolution and together will probe the
warm, inner, planet-forming region (MIRI) and cold outer parts (ALMA) of young stellar systems \citep{vandishoeck2004}.

\subsection{Galaxy assembly}

Even for well-observed local galaxies, it has proven difficult to measure 
star formation rates accurately because of the effects of absorption and 
scattering by interstellar dust (e.g., \citet{calzetti2010}). Dust 
obscuration remains a significant obstacle to accurate measures of star 
formation to z of 2 and beyond (e.g., \citet{reddy2010}), and the available 
data ({\it Spitzer}, {\it Herschel}) cannot survey deep enough to characterize it 
completely (see, e.g., Figure 4 in Elbaz et 
al. 2011). With careful calibration of the evolution of infrared spectral 
energy distributions, it has been found that photometry at 24 $\mu $m can 
measure star formation rates accurately (typical errors of 0.13 dex) out to 
z $\sim$ 2.5 (\citet{rujopakarn2013}), see Figure 7. Thus, deep 
MIRI surveys at 21 $\mu $m will be able to measure dust-embedded star 
formation rates to z $=$ 2 and down to far infrared luminosities of 3 X 
10$^{\mathrm{10}}$ L$_{\mathrm{\odot }}$. To complete 
our understanding of dust-embedded star formation at even higher redshift, z $\ge$ 2, 
accurate star formation rates in dusty galaxies will require ALMA observations. 

The reason that galaxies at z $\sim$ 1 and above have relatively similar infrared spectral
energy distributions is that their star formation, even at very high luminosities, tends to
be spread over a multi-kpc diameter region. High optical depths in the mid-infrared are
therefore rare and seldom have a substantial effect on the mid-infrared output, allowing
a standard template to produce accurate estimates of the star formation rates \citep{rujopakarn2013}. 
This situation contrasts with local very luminous galaxies, where the star formation
tends to be concentrated in regions well under a kpc in diameter centered on the
galaxy nucleus. To probe the morphologies of distant vigorously star forming galaxies requires
high resolution imaging in the mid-infrared. Fortunately, this goal can be achieved by inverting
gravitationally lensed images as shown, for example, for the huge lensed arc in Abell 370 by \citet{richard2009}.
In these cases, the resolution of MIRI will provide good reconstructed images and the MRS
IFU will make observations efficient as shown in Figure 8.

At higher redshifts, MIRI will make fundamental contributions to understanding
the mass assembly of galaxies by achieving images of unprecedented depth and
resolution matching the deepest images currently available in the near infrared (Figure 9). 
It is expected that the distribution of galaxy masses will be strongly skewed toward small
masses by z $\sim$ 5, early in the process of galaxy assembly (Figure 10). Quantifying this behavior requires reliable
determination of galaxy masses by measuring and modeling the stellar spectral energy distributions nearly out
to their peak at a restframe wavelength of 1.6 $\mu$m (rest). At z = 4, this peak has been shifted to 8 $\mu$m, where
{\it Spitzer} Infrared Array Camera (IRAC) data have been used to model galaxy masses. However, as
shown in Figure 10, the sensitivity limitations (in large part due to confusion) have kept us from
measuring masses down into the interesting range. The additional depth with MIRI
(dashed lines in Figure 10) will probe galaxy mass assembly definitively. 

\subsection{The first galaxies}

Finding the first galaxies was the original justification for JWST and 
remains its top priority. There are a number of recent theoretical
studies of what this goal might entail (e.g., \citet{pawlik2011, pawlik2013, zackrisson2011, zackrisson2012, safranek2012,  muratov2013}). These works predict a substantial range of possible
properties (e.g., \citet{zackrisson2011, safranek2012}), but generally agree that
a ``typical" first light galaxy will be of very low metallicity (by definition) and also of 
low mass and very faint, close to the
detection limits with NIRCam (the JWST Near Infrared Camera). In such cases, JWST will be able to find candidates
based on photometric redshifts, but characterizing the candidates (and even
confirming that they are first light objects, i.e., with no metals and no prior history of star formation) will be challenging. 

Therefore, systematic study of galaxies at these epochs will focus on the brighter examples. 
It is now predicted that Population III galaxies (i.e., with zero or only trace metallicity) 
may exist in significant numbers over
a broad range of redshift (see Figure 11), down
to z $\le$ 7 \citep{zackrisson2012, muratov2013}. The luminosity distance is reduced by a factor of
1.8 (or 3.1) at this redshift compared with z = 10 (or 16), corresponding to increases in brightness by 3 (or 10). 
In addition, there will be a large range of halo masses, with a corresponding range of 
galaxy masses; studies will by necessity focus on these higher mass galaxies \citep{bromm2011}. 
For example, \citet{behroozi2014} find that a single NIRCam survey field could have
hundreds of galaxies between z = 9 and 10 and with mass $>$ 10$^8$ M$_\odot$ (the upper
limit to first-light galaxy masses in many theoretical studies \citep{bromm2011}). \citet{oesch2014} report a significant number
of candidates for galaxies at z $\sim$ 10 and M $\sim$ 10$^9$ M$_\odot$. Such galaxies are
where we will definitively test the ``first light" hypothesis and use MIRI and the other
JWST instruments to study key characteristics
such as the history of star formation, the presence of
active nuclei, and the  buildup of metals. 

MIRI will play central roles in the first three of these four investigations. 
From Figure 12, MIRI colors (with NIRCam ones) can identify true Population III galaxies.
MIRI detections of H$\alpha $ (which moves into the MIRI spectral range for z $>$ 6.7) 
will give insights to the instantaneous star formation 
rates \citep{paardekooper2013}. A 30-hour integration with the MIRI medium-resoutin spectrometer on a galaxy at z = 10 would measure (4-$\sigma$) the H$\alpha$ from star formation at 
the rate of 8 M$_\odot$/yr. Candidate objects at this redshift and star formation level are
already being identified \citep{labbe2013, holwerda2014}. H$\alpha$ line profiles will also be useful to find active galactic nuclei (AGNs). 
Very low metallicity galaxies have no strong emission lines between Lyman $\alpha$ and the Balmer series
\citep{groves2008}; predictions of the strength of Ly$\alpha$ are uncertain \citep[e.g.][]{anders2003, barros2014}, with 
ndications that it may be very weak at z $>$ 7 \citep[e.g.,][]{jiang13}. Hence, H$\alpha$ with MIRI may be the most readily detectable emission
line for galaxies at z $>$ 10. At these redshifts, MIRI photometry will also be the best way to
determine the strength of Balmer breaks and hence the longer term history of star formation. 

\section{Introduction to series of papers}

This special issue of PASP contains a series of papers describing MIRI in full. The papers  
deal with each of the main science capabilities in turn, including imaging, prism 
spectroscopy, coronagraphy and integral field spectroscopy. 
Key to achieving the science goals is the understanding and 
optimisation of the performance of the three detector arrays in MIRI and 
of their control electronics.The operating characteristics of the 
detectors themselves and of their readouts and control electronics are therefore described 
separately. The paper on sensitivity 
then combines our understanding of the performance of the entire electro-optic 
detection chain with the expected observatory environment, into an estimation of the 
sensitivity that will be achieved in an observation using a single spacecraft pointing 
on-orbit. Finally, a snapshot of the still developing operating model and calibration pipeline 
for MIRI observing is presented, where the plans for calibrating, mapping and 
background subtraction are illustrated.

The specific papers are:


\noindent
II: ``Design and Build" 
\citep{wright2014} is an overall introduction to the instrument system and its thermal, mechanical, electronic 
and software  interfaces with the JWST  spacecraft. It also describes subsystems common to all the
 individual instrument capabilites (e.g., calibration sources). 

\noindent
III: ``MIRIM, the MIRI imager" \citep{bouchet2014} gives an overview of the imaging module, which
also supports the prism spectrometer and coronagraphic capabilities. It describes the overall optical design,
and gives the details for the imaging band-defining filters. It also has an overview of the test campaign for
the imager and its results.

\noindent
IV: ``The MIRI Low Resolution Spectrograph" \citep{kendrew2014} describes the double prism in the
imager filter wheel that provides spectra with resolution of $\lambda$/$\Delta\lambda$ $\sim$ 100, 
used either with a single slit (for spectra of very faint objects) or in a slitless mode (primarily envisioned 
for planetary transit measurements).

\noindent
V: "The Predicted Performance of the MIRI Coronagraphs," \citep{boccaletti2014} describes another capability 
included in the MIRI imager. There are four coronagraphs, three based on quarter wave phase plates and one
a classical Lyot design. The paper describes the principles of operation, the implementation in MIRI, and then models
the expected performance in detail, taking account of the current best estimates for the telescope wavefront
error and guiding accuracy.

\noindent
VI: ``The Medium-Resolution Spectrometer" \citep{wells2014} discusses the principles behind the optical design of the general purpose MIRI spectrometer, followed by a description of its construction and ground-test results. It concludes with 
a number of areas required for effective use of the instrument, such as correction for fringing in the detector
arrays, generating calibration data, and corrections for stray light. 

\noindent
VII:``The MIRI Detectors" \citep{rieke2014} describes the general principles of operation of the
Si:As IBC detectors in the instrument and their basic performance properties. It follows with more detailed theoretical analyses of : 1.) the trend of response with bias voltage; 2.) the detector/readout nonlinearity; 3.)  latent images; and 4.) the
cross-like image artifact in the 5 - 8 $\mu$m spectral range. 

\noindent
VIII: ``The MIRI Focal Plane System" \citep{ressler2014} shows how the detector array readout circuits work,
how they are controlled, and how their signals are digitized and formatted. It discusses the options for subarrays. 
Finally, it includes a listing of the primary anomalies in the photometric performance of the detectors that need
to be corrected in the data pipline.

\noindent
IX: ``Predicted Sensitivity" \citep{glasse2014} combines the test results on the flight instrument into our best projection of the signal to noise it will achieve on orbit.

\noindent
X: ``Operations and Data Reduction" \citep{gordon2014} shows how MIRI observations will be planned 
using Observation Templates, one template for each of the primary instrument modes (imaging, coronagraphy, 
low resolution spectroscopy, medium resolution spectroscopy). It then outlines the approach being developed
for data reduction and closes with some samples of potential MIRI observations. 


\section{Acknowledgements}
The work presented is the effort of the entire MIRI team and the
enthusiasm within the MIRI partnership is a significant factor in its
success. MIRI draws on the scientific and technical expertise of the
following organisations: Ames Research Center, USA; Airbus Defence and
Space, UK; CEA-Irfu, Saclay, France; Centre Spatial de Li\'{e}ge,
Belgium; Consejo Superior de Investigaciones Cient\'{\i}ficas, Spain;
Carl Zeiss Optronics, Germany; Chalmers University of Technology,
Sweden; Danish Space Research Institute, Denmark; Dublin Institute for
Advanced Studies, Ireland; European Space Agency, Netherlands; ETCA,
Belgium; ETH Zurich, Switzerland; Goddard Space Flight Center, USA;
Institute d'Astrophysique Spatiale, France; Instituto Nacional de
T\'{e}cnica Aeroespacial, Spain; Institute for Astronomy, Edinburgh,
UK; Jet Propulsion Laboratory, USA; Laboratoire d'Astrophysique de
Marseille (LAM), France; Leiden University, Netherlands; Lockheed
Advanced Technology Center (USA); NOVA Opt-IR group at Dwingeloo,
Netherlands; Northrop Grumman, USA; Max-Planck Institut f\H{u}r
Astronomie (MPIA), Heidelberg, Germany; Laboratoire d'Etudes Spatiales et 
d'Instrumentation en Astrophysique (LESIA), France;
Paul Scherrer Institut, Switzerland; Raytheon Vision Systems, USA;
RUAG Aerospace, Switzerland; Rutherford Appleton Laboratory (RAL
Space), UK; Space Telescope Science Institute, USA;
Toegepast-Natuurwetenschappelijk Onderzoek (TNO-TPD), Netherlands; UK
Astronomy Technology Centre, UK; University College London, UK;
University of Amsterdam, Netherlands; University of Arizona, USA;
University of Bern, Switzerland; University of Cardiff, UK; University
of Cologne, Germany; University of Ghent; University of Groningen,
Netherlands; University of Leicester, UK; University of Leuven,
Belgium; University of Stockholm, Sweden; Utah State University, USA.
A portion of this work was carried out at the Jet Propulsion Laboratory, 
California Institute of Technology, under a contract with the National 
Aeronautics and Space Administration.

We would like to thank the following National and International
Funding Agencies for their support of the MIRI development: NASA; ESA;
Belgian Science Policy Office; Centre Nationale D'Etudes Spatiales (CNES);
Danish National Space Centre; Deutsches Zentrum fur Luft-und Raumfahrt
(DLR); Enterprise Ireland; Ministerio De Economi{\'a} y Competividad;
Netherlands Research School for Astronomy (NOVA); Netherlands
Organisation for Scientific Research (NWO); Science and Technology Facilities
Council; Swiss Space Office; Swedish National Space Board; UK Space
Agency.

We take this opportunity to thank the ESA JWST Project team and the
NASA Goddard ISIM team for their capable technical support in the
development of MIRI, its delivery and successful integration.

We would like to thank Peter Jakobsen for his leadership of the European interests in 
JWST as the partnership in the JWST project was established, and in particular for 
his strong support of MIRI over many years. 
We also thank Fred Lahuis and Klaus Pontippidan for help with Figures 5 and 6, respectively. 
We also thank Lisa May Walker for a critical reading of a draft of the paper.

\begin{deluxetable}{ccccc}
\tabletypesize{\footnotesize}
\tablecolumns{5}
\tablewidth{0pt}
\tablecaption{ MIRI Measurement Capabilities}
\tablehead{\colhead{name}           &
	 \colhead{FOV}             &
           \colhead{Wavelength}             &
	 \colhead{Spectral Properties}     &
            \colhead{Reference}             \\
            \colhead{}                      &
            \colhead{}                      &
            \colhead{Range ($\mu$m)}   &
            \colhead{}                     &
            \colhead{(this volume)}         \\
  }
\startdata
Diffraction-limited Imaging  &  $74'' \times 113''$  & 5.6 - 25.5  & 9 bands &  Paper III\\
Low Res. Spectroscopy & 0\farcs51 $\times$ 4\farcs7 slit  & 5 - 12  & $\lambda/\Delta\lambda \sim 100$ & Paper IV\\
Slitless Spectroscopy & 7\farcs9 wide  &  5 - 12  & $\lambda/\Delta\lambda \sim 100$  &  Papers IV \& VIII  \\
Phase Mask Coronagraphy &  24$''$ $\times$ 24$''$ & 10.65 - 15.5  & 3 bands & Paper V \\
Lyot Coronagraphy & 30$''$ $\times$ 30$''$ &  23   & one band  & Paper V  \\
Medium Res. Spectroscopy &  3\farcs44 $\times$ 3\farcs64 ~IFU$^a$  &  4.9 - 28.8 & $\lambda/\Delta \lambda$ $\sim$ 1500 - 3500  &  Paper VI \\
\enddata
\tablenotetext{a}{Four integral field units (IFUs); the listed field is common to all, while the longer wavelength units have larger fields up to $7'' \times 7''$}
\end{deluxetable}

\begin{deluxetable}{cccccc}
\tabletypesize{\footnotesize}
\tablecolumns{5}
\tablewidth{0pt}
\tablecaption{ Imager and Coronagraph Properties}
\tablehead{\colhead{name}           &
	 \colhead{FOV$^a$ }             &
           \colhead{$\lambda_0$}             &
	 \colhead{$\lambda$/$\Delta\lambda$}     &
            \colhead{10-$\sigma$ in }   &
            \colhead{$\lambda$/D }                         \\
          \colhead{}                                                &
          \colhead{(arcsec)}                                  &
          \colhead{($\mu$m)}                               &
          \colhead{}                                              &
          \colhead{10,000s$^d$}                        &
          \colhead{(arcsec)}                                 
  }
\startdata
F560W  &  74 $\times$ 113  &5.6 & 5 &  0.17 $\mu$Jy  &  0.19  \\
F770W & 74 $\times$ 113  & 7.7 & 3.5 & 0.27 $\mu$Jy  &  0.26  \\
F1000W &  74 $\times$ 113 & 10 & 5 &  0.60 $\mu$Jy   &  0.34  \\
F1130W  & 74 $\times$ 113 & 11.3  & 16 & 1.48 $\mu$Jy  &  0.39  \\
F1280W  & 74 $\times$ 113  & 12.8  &  5  & 0.94 - 1.05 $\mu$Jy &  0.44  \\
F1500W  &  74 $\times$ 113  & 15  &   5 & 1.5 - 2.0 $\mu$Jy  &  0.52  \\   
F1800W  & 74 $\times$ 113 &  18  &  6 & 3.7 - 5.3 $\mu$Jy &  0.62   \\
F2100W  & 74 $\times$ 113 & 21  & 4  & 7.5 - 10.5 $\mu$Jy  &  0.72  \\
F2550W  & 74 $\times$ 113 &  25.5 & 6 & 27 - 36 $\mu$Jy  &  0.88  \\ 
F1065C$^b$  & 24 $\times$ 24 & 10.65  & 20  &  &  0.37  \\
F1140C$^b$  & 24 $\times$ 24  & 11.40  &  20 &  & 0.39  \\
F1550C$^b$  & 24 $\times$ 24 &  15.5   &  20 &  & 0.53  \\
F2300C$^c$  & 30 $\times$ 30 &  23     &   5  & & 0.79  \\
\enddata
\tablenotetext{a}{All imager functions have 0\farcs11 projected pixels}
\tablenotetext{b}{Four quadrant phase mask coronagraph}
\tablenotetext{c}{~Lyot coronagraph}
\tablenotetext{d}{The range of detection limits at the long wavelengths reflects the uncertainty in the estimates of the observatory emission level in these bands.}
\end{deluxetable}

\clearpage

\begin{deluxetable}{ccccc}
\tabletypesize{\footnotesize}
\tablecolumns{5}
\tablewidth{0pt}
\tablecaption{Spectrometer Properties}
\tablehead{\colhead{name}           &
           \colhead{FOV}             &
	 \colhead{sub-band}               &    
	 \colhead{$\lambda$ range}              &
 	\colhead{$\lambda/\Delta\lambda$}             \\
          \colhead{}                                     &
          \colhead{slice width, pixel size}    &
         \colhead{}                                     &
         \colhead{ ($\mu$m)}                                    &
          \colhead{}                                                 \\
          \colhead{}                                 &
          \colhead{(arcsec)}                    &
          \colhead{}                                &
         \colhead{}                                 &
         \colhead{}
  }
\startdata
     LRS      &  0.51 $\times$ 4.7  &    &  5 - $\sim$ 14  &  $\sim$ 100$^a$   \\
                 &      -----, 0.11                          &    &                         &                              \\
      MRS      &                            &   A   &  4.87 - 5.82   &                      \\
Channel 1    & 3.0 $\times$ 3.9 & B   &   5.62 - 6.73  &  2450 - 3710  \\
                  &      0.176, 0.196        & C  &    6.49 - 7.76  &               \\
       MRS       &                            & A  & 7.45 - 8.90   &             \\
Channel 2   & 3.5 $\times$ 4.4  & B  & 8.61 - 10.28 & 2480 - 3690 \\
                   &       0.277, 0.196        &  C & 9.94 - 11.87&                   \\   
      MRS        &                             & A & 11.47 - 13.67&                 \\
Channel 3   & 5.2 $\times$ 6.2  & B &  13.25 - 15.80 & 2510 - 3730  \\
                   &        0.387, 0.245         & C  & 15.30 - 18.24 &               \\ 
      MRS         &                              & A  & 17.54 - 21.10 &               \\
Channel 4   & 6.7 $\times$ 7.7  & B  & 20.44 - 24.72 & 2070 - 2490  \\
                   &      0.645, 0.273            & C   & 23.84 - 28.82  &               \\
\enddata
\tablenotetext{a}{at 7.5$\mu$m. The detection limit at this wavelength is estimated to be 3 $\mu$Jy, 10-$\sigma$, 10,000 seconds.}
\end{deluxetable}

\clearpage

\begin{figure}[htbp]
\centerline{\includegraphics[width=6.0in]{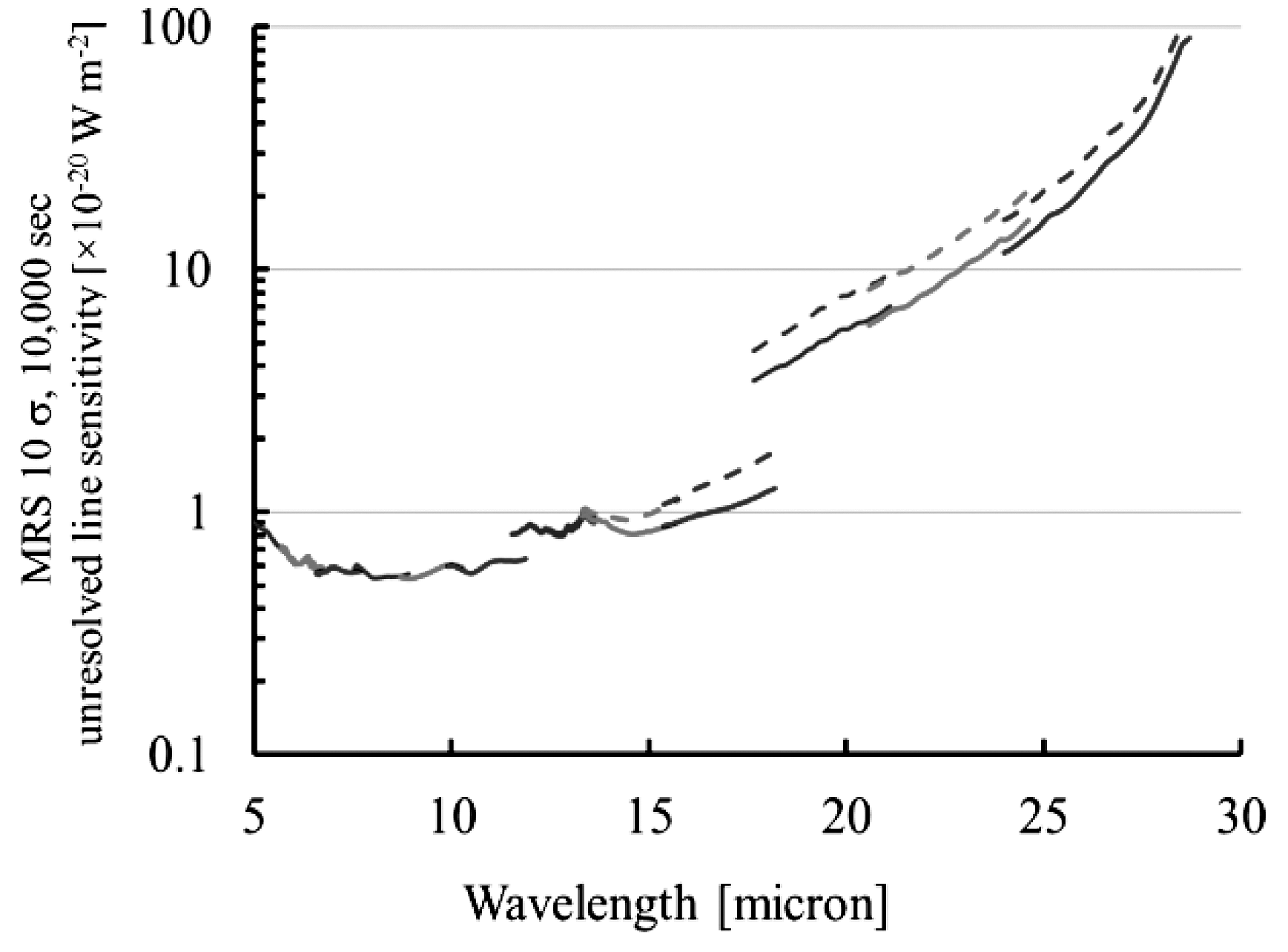}}
\caption{Estimated detection limits (10-$\sigma$ in 10,000 seconds of integration) for an unresolved spectral line from a point source, with the medium resolution spectrometer (MRS). The solid and dashed lines show estimates for two different possible
levels of observatory emission as discussed in Paper IX. }
\label{fig:fig1}
\end{figure}

\clearpage

\begin{figure}[htbp]
\centerline{\includegraphics[width=6.0in]{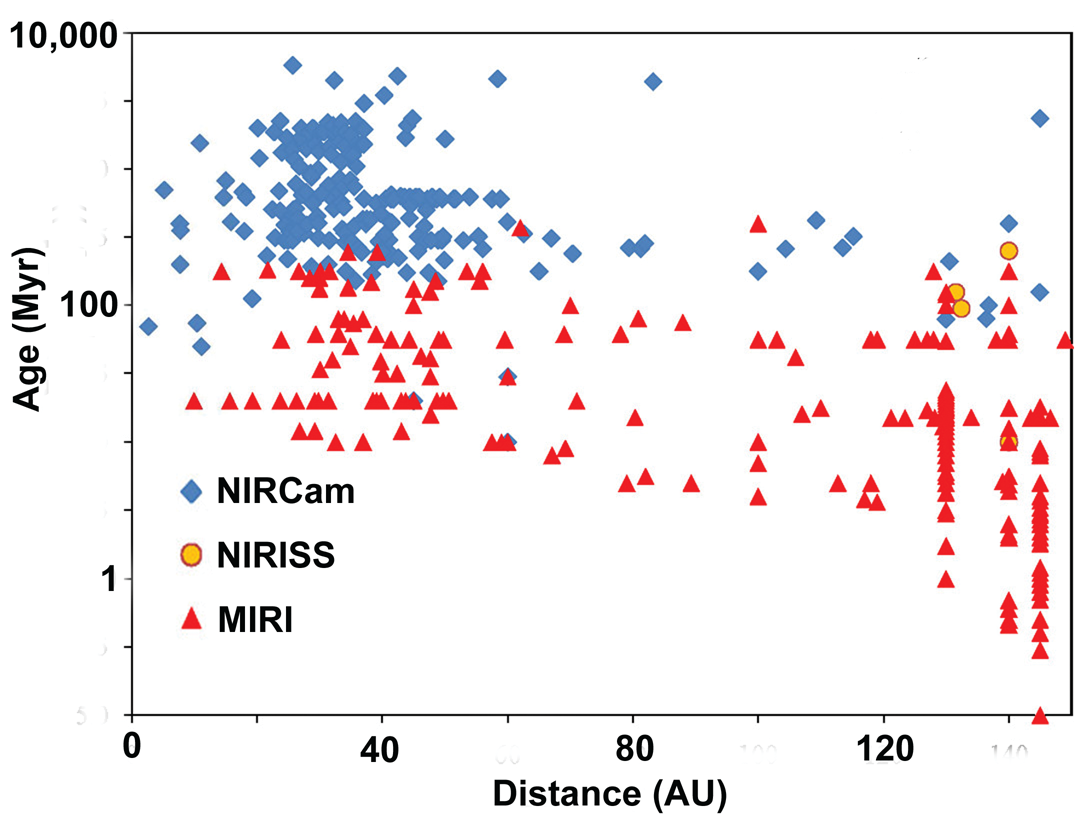}}
\caption{Detectability of young planets as a function of age and distance from their stars, based on 
a Monte Carlo analysis of high-contrast (coronagraphic) JWST  imaging. The investigation was
centered on the stars most likely to have detectable planets (e.g., nearby, and young or of 
very low mass) and planet fluxes were predicted from the CONDO3/DUSTY models \citep{baraffe2003}. The symbols are 
coded according to the instrument with the highest likelihood of achieving a detection;
MIRI is the instrument of choice for planets younger than 200 Myr, NIRCam for most cases for older planets, and the Near-IR Imager and 
Slitless Spectrograph (NIRISS) for those closest to their stars. More importantly, many of the simulated planets would be detected in 
multiple ways, allowing derivation of many of their intrinsic properties \citep{beichman2010}. }
\label{fig:fig1}
\end{figure}

\clearpage

\begin{figure}[htbp]
\centerline{\includegraphics[width=4.0in]{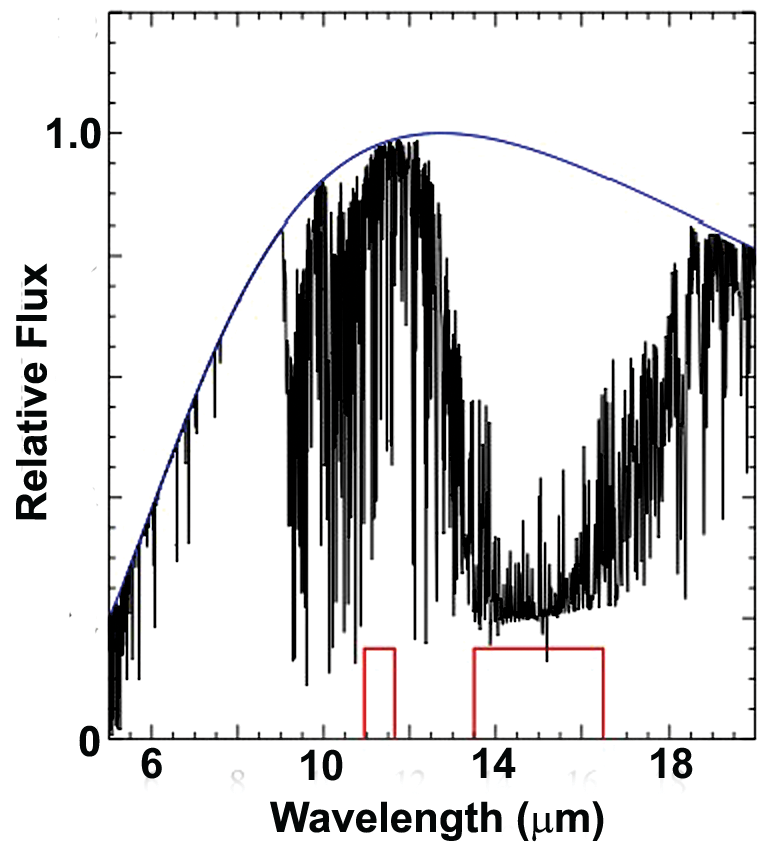}}
\caption{Simulated spectrum of a super-Earth and two MIRI photometric bands. The spectral continuum is in blue and the effects of absorption features in black. The MIRI imager spectral bands at 11.3 and 15 $\mu$m are shown in red along the x-axis \citep{deming2009}. 
The depth of the CO$_2$ absorption indicates whether the planetary atmosphere is hydrogen poor (shown here) or
hydrogen rich (in which case there is virtually no absorption) \citep{miller2009}. It therefore allows a critical test of whether the 
planet is habitable in any conventional sense. 
Figure 4 shows the number of systems that could be measured in this way\citep{deming2009}. }
\label{fig:fig2}
\end{figure}

\clearpage

\begin{figure}[htbp]
\centerline{\includegraphics[width=5.0in]{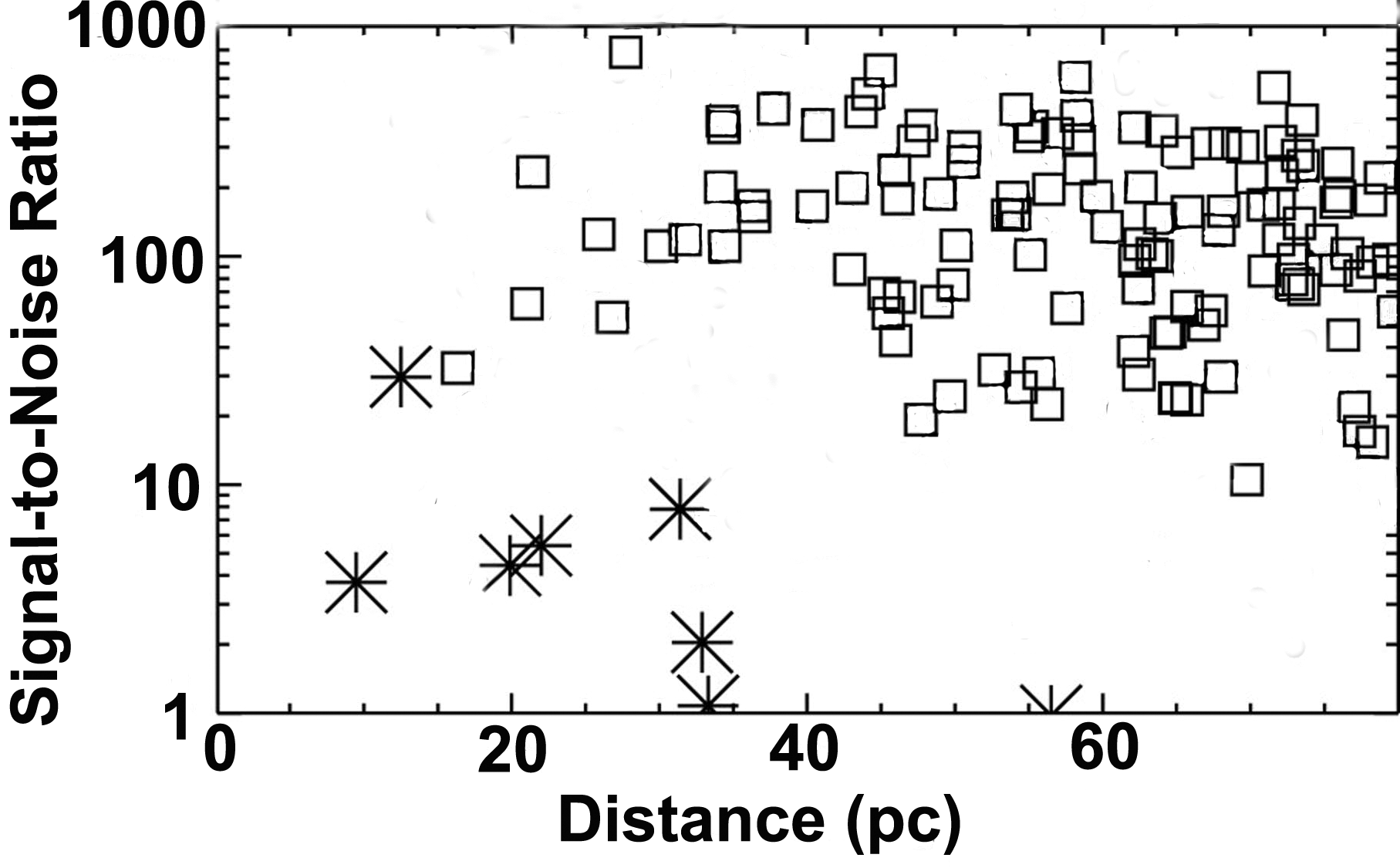}}
\caption{Signal to noise ratios for simulated TESS planets, hypothesized to be measured by MIRI eclipse 
photometry at 15 $\mu$m. The star symbols represent super-Earths in the habitable zones of their stars and apply to detection of CO$_2$ as in Figure 3, with signal to noise assuming all available transits are measured with JWST. The open squares are planets with radii between 3 \& 5 Earth radii (i.e., Neptunes), and the signal to noise is for continuum radiation at 15 $\mu$m. \citep{deming2009}. }
\label{fig:fig3}
\end{figure}

\clearpage

\begin{figure}[htbp]
\centerline{\includegraphics[width=6.0in]{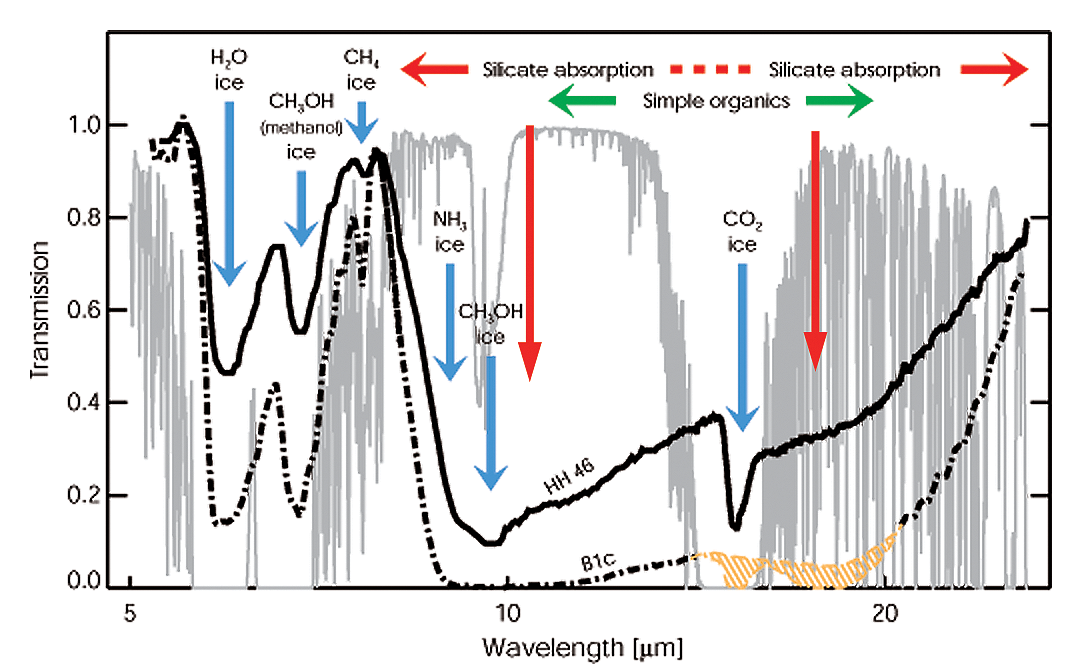}}
\caption{Windows for probing cold cloud cores. The grey tracing is the transmission of the terrestrial atomosphere (1mm water vapor). The two {\it Spitzer} IRS low-resolution spectra are of lightly (HH 46), and heavily (the B1c protostar in Persius (Boogert et al. 2008)) obscured sources. For the latter, a spline area interpolation has been used (hashed green area) to show what might 
be expected for the spectral region not observed with IRS. The positions of important spectral features are marked. The 10 and 18 $\mu$m silicate absorptions are indicated by the downward pointing red arrows. They leave windows into the cloud core at 5 - 8$\mu$m and 14 - 17 $\mu$m, but these ranges are blocked by the terrestrial atmosphere. }
\label{fig:fig4}
\end{figure}

\clearpage

\begin{figure}[htbp]
\centerline{\includegraphics[width=6.0in]{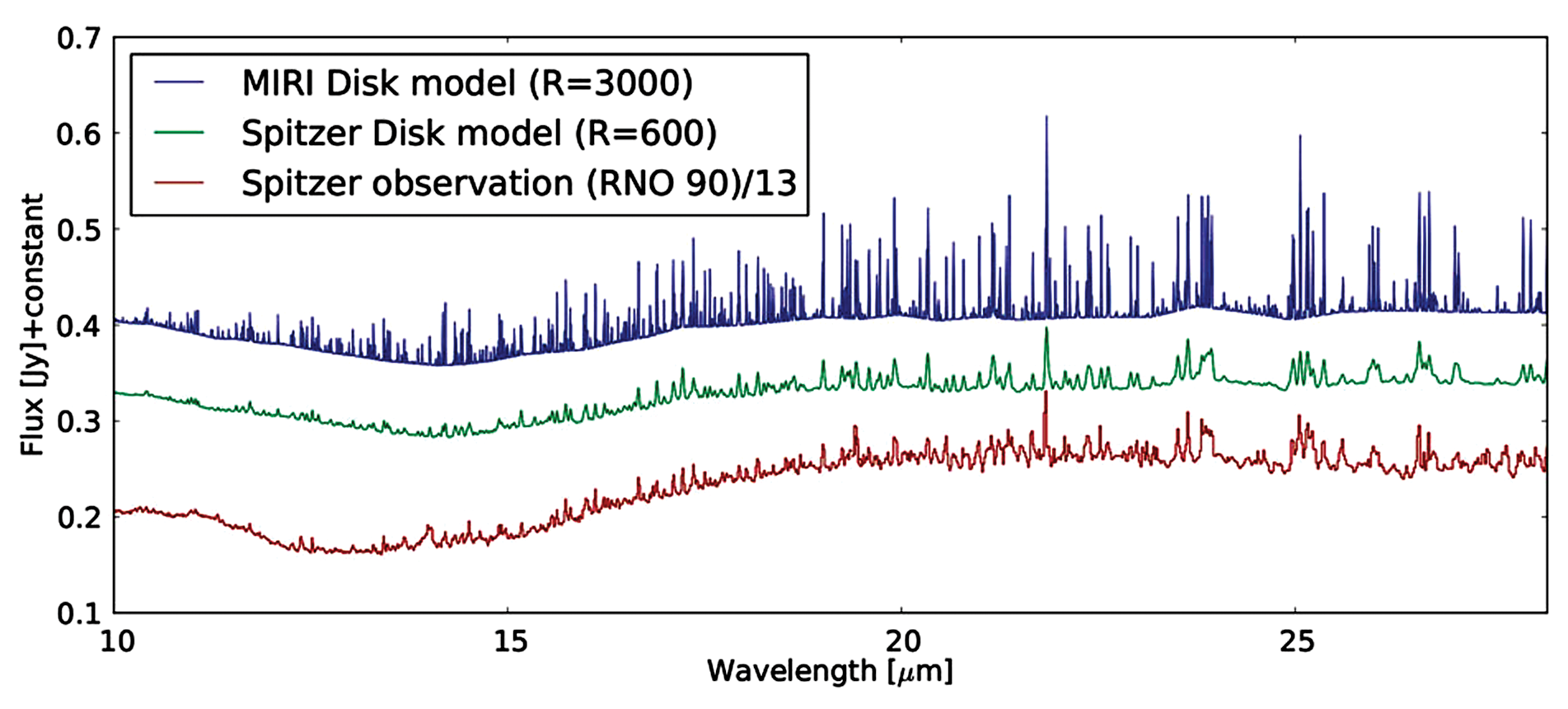}}
\caption{The lowest spectrum is an observation with the IRS on {\it Spitzer} of the molecular forest in the protoplanetary disk
around RNO 90 (Pontoppidan et al. 2010); the spectrum above it is a two-dimensional radiative transfer model of water vapor 
adjusted to the same resolution (R = 600). The upper spectrum is the same model but at the MIRI resolving power, R $\sim$ 3000.
Although the spectrum of RNO 90 is dominated by H$_2$O transitions, the forest of lines should also contain contributions from OH, and from ro-vibrational bands of simple organic molecules such as CO$_2$, HCN, and C$_2$H$_2$.  
Illustration from K. M. Pontoppidan, STScI, 
https~://blogs.stsci.edu/newsletter/2013/03/27/new-window-into-planet-formation-with-webb
}
\label{fig:fig4}
\end{figure}

\clearpage

\begin{figure}[htbp]
\centerline{\includegraphics[width=6.0in]{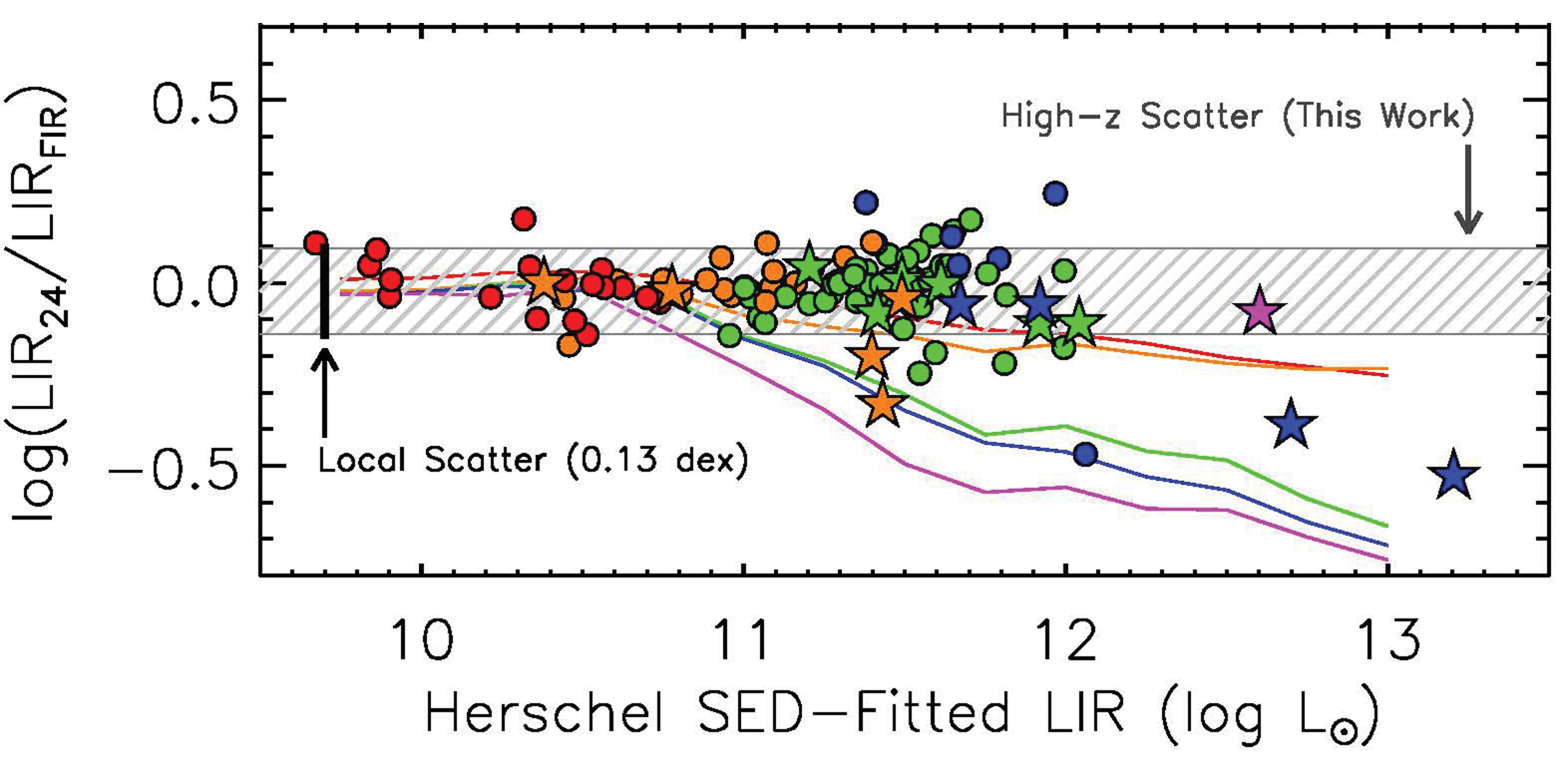}}
\caption{Comparison of L(TIR) (the total infrared luminosity) estimated from 24 $\mu$m measurements only with that from
integrating spectral energy distributions (SEDs) fitted to {\it Herschel} SPIRE (Spectral and Photometric Imaging Receiver) 
photometry  \citep{rujopakarn2013}. This figure 
refers to purely star-forming galaxies only and to apply the results requires that strong active nuclei
be eliminated from the sample, e.g., through deep X-ray imaging and identification of power-law objects in the mid-IR. 
The ratio of the two determinations (LIR$_{24}$ and LIR$_{FIR}$) has a scatter
of 0.12 dex, similar to the scatter in determining L(TIR) from 24 $\mu$m measurements locally. The
individual points are color-coded to redshift, red (z $<$ 0.3), orange (0.3 $<$ z $<$ 0.6), green
(0.6 $<$ z $<$ 1.0, blue (1.0 $<$ z $<$ 2.0) and purple (z $>$ 2.0). The lines, color coded
similarly, show the ratios that would result if the high-redshift galaxies had infrared SEDs similar
to those of local ones of the same luminosities, demonstrating the importance of accounting for the
SED evolution in this result. About 6\% of high redshift star-forming galaxies appear to be
compact like local ones, and for them determination of the star formation rate from 24 $\mu$m will
have larger errors.}
\label{fig:fig5}
\end{figure}

\clearpage

\begin{figure}[htbp]
\centerline{\includegraphics[width=6.0in]{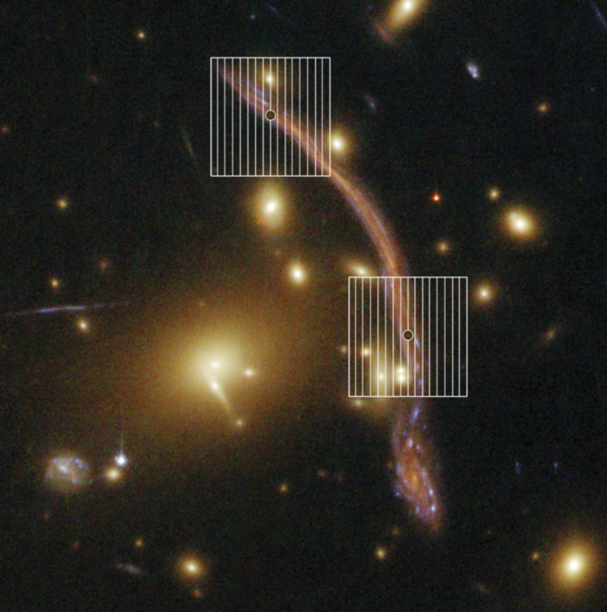}}
\caption{The giant arc lensed by the Frontier Fields cluster Abell 370 and imaging a background galaxy at z = 0.725 \citep{richard2009}. Superimposed on the arc are two MIRI IFU fields for Channel 3 (11.47 - 18.24 $\mu$m). The vertical lines delineate the IFU slices, while the circle in the middle of each field is $\lambda$/D at the middle wavelength of Channel 3 (15.5 $\mu$m). In this example, the MIRI reconstructed image could reveal the distribution of aromatic features, characterize the 
radiation field through the relatively extinction-free mid-infrared fine structure lines of [ArII] and [ArIII] and [NeII] and [NeIII], 
and identify continuum warm spots marking heating by embedded massive stars. }
\label{fig:fig6}
\end{figure}

\clearpage

\begin{figure}[htbp]
\centerline{\includegraphics[width=6.0in]{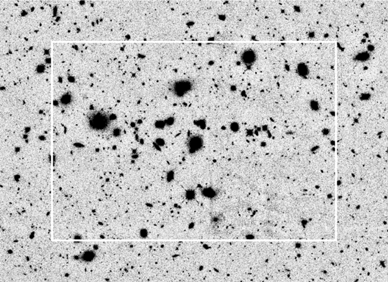}}
\caption{MIRI images at 5.6 and 7.7 $\mu$m will reach the depth and resolution to match very deep near infrared images such as the one above at 1.6 $\mu$m from CANDELS \citep{koekemoer2011}. The resolution of this image (0\farcs2) matches closely
that of MIRI at 5.6 $\mu$m, and the white box shows the FOV of the MIRI imager ($74'' \times 113''$). At this wavelength the MIRI detection limits will be within a magnitude (AB) of those in this image, allowing measurement of virtually all the galaxies at reasonably 
high redshift \citep{caputi2011}.}
\label{fig:fig6}
\end{figure}

\clearpage

\begin{figure}[htbp]
\centerline{\includegraphics[width=6.0in]{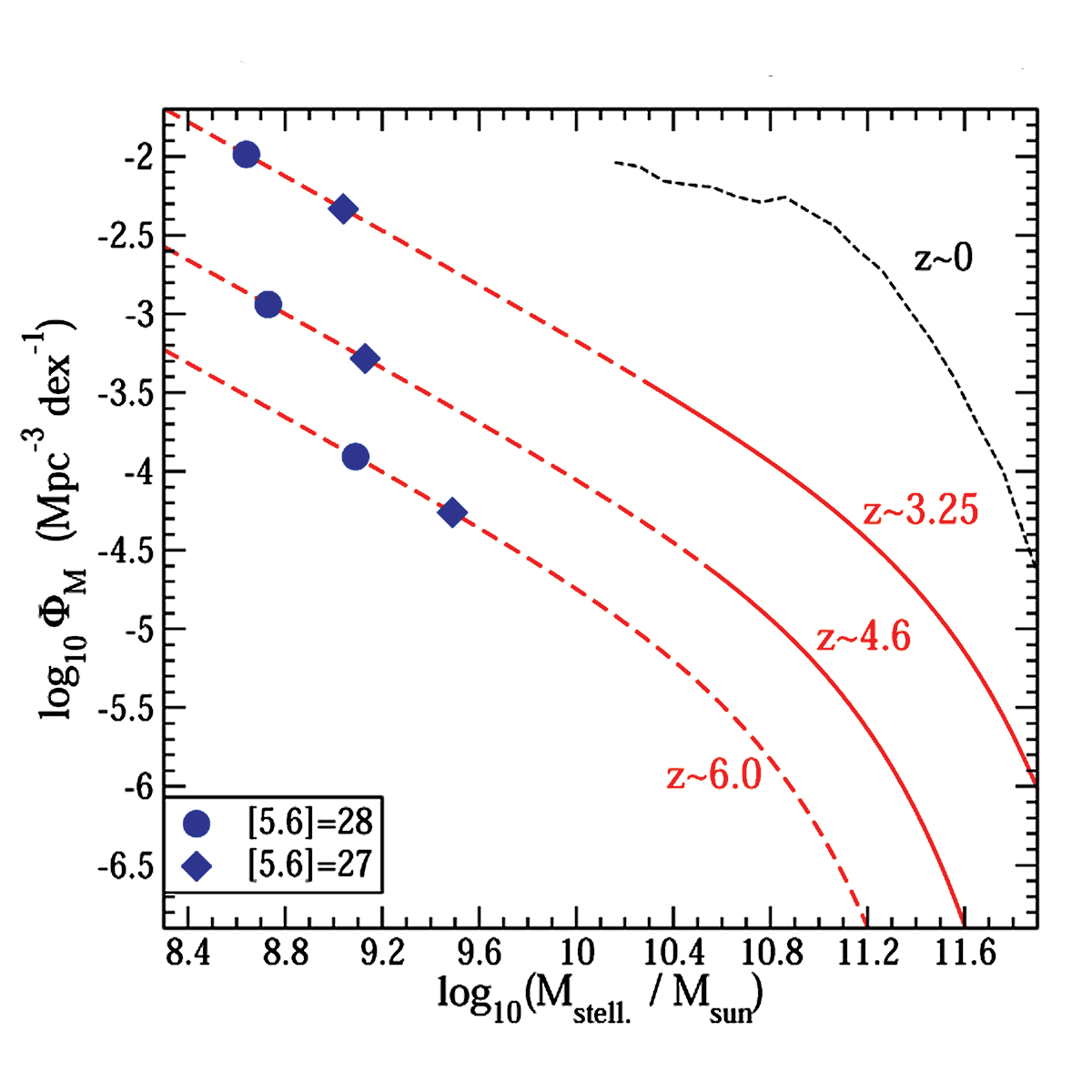}}
\caption{Theoretical predictions of the evolution of the galaxy mass function with redshift from \citet{caputi2011}, according to the models of \citet{choi2010}. 
The solid parts indicate how far down in mass it is possible to go with the deepest IRAC images \citep{caputi2011}. The additional depth with MIRI is indicated by the blue diamonds and circles indicating completeness limits corresponding to AB magnitudes
respectively of 27 and 28 at 5.6 $\mu$m, which can be reached (4-$\sigma$) in 3 and 20 hours, respectively. These measurements 
will allow exploring the dashed regions where the predicted changes with redshift should be readily apparent.}
\label{fig:fig7}
\end{figure}

\clearpage

\begin{figure}[htbp]
\centerline{\includegraphics[width=6.0in]{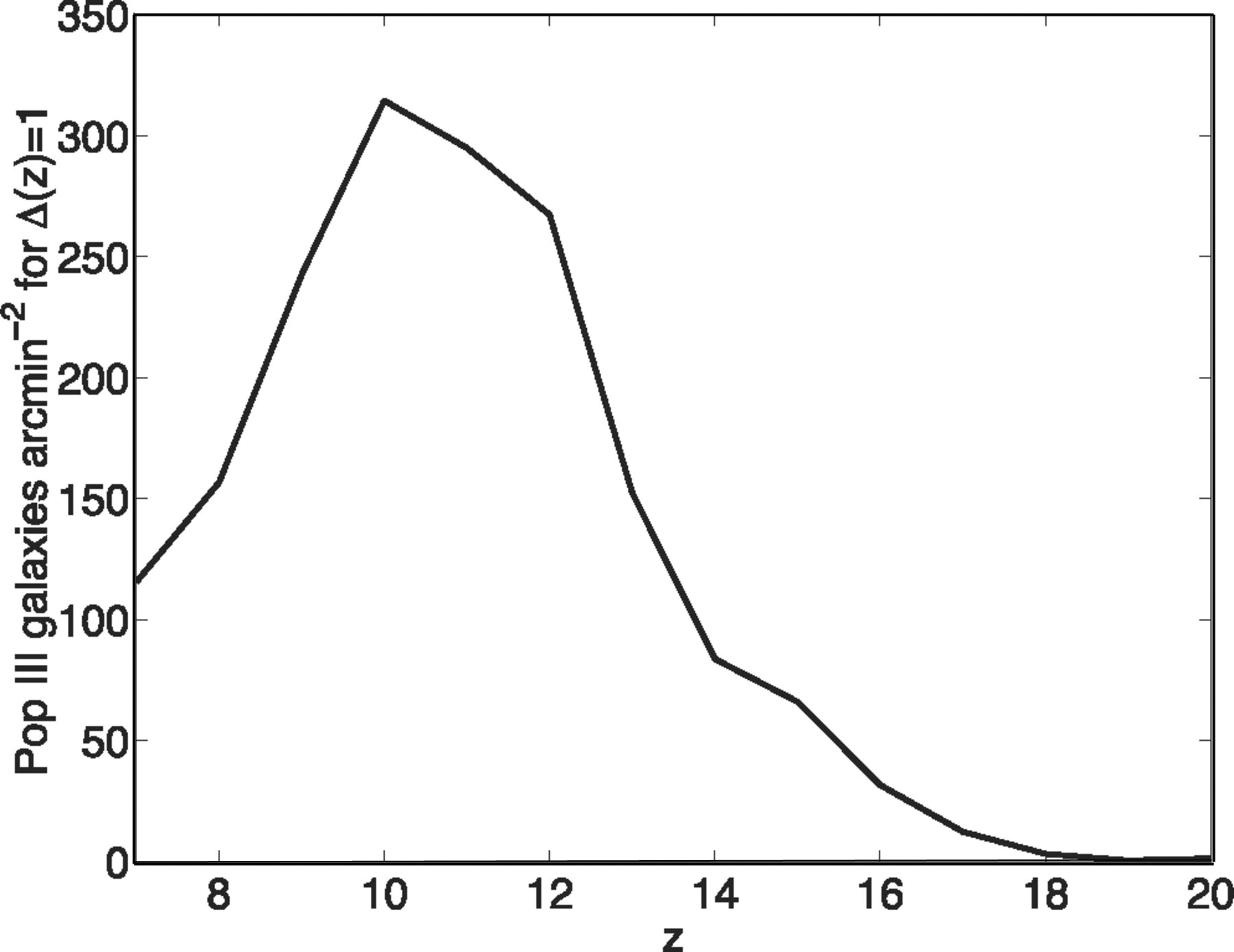}}
\caption{The predicted number of pop III galaxies per square arcmin 
and unit redshift in an unlensed field \citep{zackrisson2012}. The numbers are based on pop III halo catalogs generated as described 
by \citet{trenti2009}, i.e. by calculating the evolution from a synthetic $\Lambda$ CDM model at Z = 199. 
Halo masses are converted to fluxes as in \citet{zackrisson2011}.}
\label{fig:fig8}
\end{figure}

\clearpage

\begin{figure}[htbp]
\centerline{\includegraphics[width=6.0in,height=4.7in]{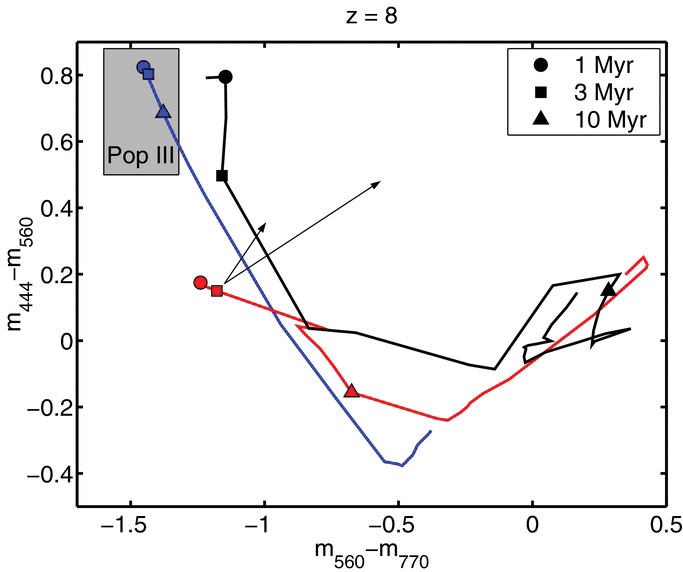}}
\caption{JWST photometric signatures of Pop III galaxies at z = 8 \citep{zackrisson2011}. 
The lines are for models of instantaneous bursts of star formation, with symbols indicating ages as  in the inset. Blue is for Pop III.2, red for Pop II and black for Pop I. The arrows show how the colors of a 3 Myr old Pop II galaxy would be affected by LMC-type (short arrow) or Calzetti (long arrow) extinction, assuming $E(B-V)$ = 0.25. The Pop III galaxies will lose their unique color
identities after some time of order a few 10$^7$ years \citep{zackrisson2011}. For more details, see 
\citet{zackrisson2011}.}
\label{fig:fig9}
\end{figure}

\end{document}